\documentclass[journal, 10pt]{IEEEtran}

\usepackage{subfigure}
\usepackage{pdfpages}
\usepackage{diagbox}
\usepackage{colortbl}
\definecolor{mygray}{gray}{0.85}
\usepackage{array}
\usepackage{caption}
\usepackage[linesnumbered,boxed,ruled,commentsnumbered]{algorithm2e}
        \usepackage{setspace}

 \usepackage{eqparbox}

\usepackage{amsmath,bm}
\usepackage{cite}
\usepackage{amsmath,amssymb,amsfonts}
\usepackage{algorithmic}
\usepackage{amsthm}
\usepackage{graphicx}
\usepackage{textcomp}
\usepackage{xcolor}

\usepackage{hyperref}
\hypersetup{%
  bookmarks=true
}

\usepackage{cite}
\usepackage{amsmath,amssymb,amsfonts}
\usepackage{algorithmic}
\usepackage{graphicx}
\usepackage{textcomp}
\usepackage{xcolor}
\usepackage{subfigure}
\usepackage{amsthm}
\usepackage{mathrsfs}

\usepackage{pdfpages}
\usepackage{colortbl}
\definecolor{mygray}{gray}{0.85}
\usepackage{array}
\usepackage{caption}
\usepackage{setspace}
\usepackage{diagbox}
\definecolor{mygray}{rgb}{0.96,0.99,0.96}
\definecolor{mypink}{rgb}{.99,.93,.85}
\definecolor{mycyan}{cmyk}{.32,0.04,0,0}
\usepackage[linesnumbered,boxed,ruled,commentsnumbered]{algorithm2e}
\usepackage{xcolor}  
\usepackage{colortbl,booktabs}
\usepackage{multirow}
\usepackage[ruled,linesnumbered]{algorithm2e}

{}
{}

\newcolumntype{I}{!{\vrule width 2.25pt}}
\newlength\savedwidth

\newlength\savewidth

\begin{document}
%
\title{Reconfigurable Intelligent Computational Surfaces: When Wave Propagation Control Meets Computing
}

\author{Bo~Yang, Xuelin~Cao, Jindan Xu, Chongwen~Huang, George C. Alexandropoulos, Linglong Dai,  M\'erouane Debbah, H. Vincent Poor, and Chau Yuen

 \thanks{B. Yang, X. Cao, J. Xu, and C. Yuen are with the Engineering Product Development Pillar, Singapore University of Technology and Design, Singapore 487372 (e-mail: bo$\_$yang, xuelin$\_$cao, jindan$\_$xu, yuenchau@sutd.edu.sg).}
 
\thanks{C. Huang is with the College of Information Science and Electronic Engineering, Zhejiang Provincial Key Lab of information processing, communication and networking, Zhejiang University, Hangzhou, 310007, China, and also with International Joint Innovation Center, Zhejiang University, Haining 314400, China (e-mail: chongwenhuang@zju.edu.cn).}

\thanks{G. C. Alexandropoulos is with the Department of Informatics and Telecommunications, National and Kapodistrian University of Athens, 15784 Athens, Greece and also with with the Technology Innovation Institute, 9639 Masdar City, Abu Dhabi, United Arab Emirates (e-mail: alexandg@di.uoa.gr).}

\thanks{L. Dai is with Beijing National Research Center for Information Science and Technology (BNRist) as well as the Department of Electronic Engineering, Tsinghua University, Beijing 100084, China (e-mails: daill@tsinghua.edu.cn).}

\thanks{M. Debbah is with the Technology Innovation Institute, 9639 Masdar City, Abu Dhabi, United Arab Emirates (email: merouane.debbah@tii.ae) and also with CentraleSupelec, University Paris-Saclay, 91192 Gif-sur-Yvette, France.}

\thanks{H. V. Poor is with Department of Electrical and Computer Engineering, Princeton University, NJ, USA (e-mail: poor@princeton.edu).}
}


\maketitle

\begin{abstract}

The envisioned sixth-generation (6G) of wireless networks will involve an intelligent integration of communications and computing, thereby meeting the urgent demands of diverse applications. To realize the concept of the smart radio environment, reconfigurable intelligent surfaces (RISs) are a promising technology for offering programmable propagation of impinging electromagnetic signals via external control. However, the purely reflective nature of conventional RISs induces significant challenges in supporting computation-based applications, e.g., wave-based calculation and signal processing. To fulfil future communication and computing requirements, new materials are needed to complement the existing technologies of metasurfaces, enabling further diversification of electronics and their applications. In this event, we introduce the concept of reconfigurable intelligent computational surface (RICS), which is composed of two reconfigurable multifunctional layers: the `reconfigurable beamforming layer' which is responsible for tunable signal reflection, absorption, and refraction, and the `intelligence computation layer' that concentrates on metamaterials-based computing. By exploring the recent trends on computational metamaterials, RICSs have the potential to make joint communication and computation a reality. We further demonstrate two typical applications of RICSs for performing wireless spectrum sensing and secrecy signal processing. Future research challenges arising from the design and operation of RICSs are finally highlighted.
\end{abstract}


\IEEEpeerreviewmaketitle

\section{Introduction}
As a key enabler for building smart wireless environments, metamaterials, sometimes known as metasurfaces, are engineered materials with promising artificial properties that are not exhibited by natural materials. Recent advances in the design of such materials offer exciting opportunities for unprecedented control and manipulation of electromagnetic (EM) properties, thereby promoting the emergence of reconfigurable/programmable metasurfaces. In such a context, reconfigurable intelligent surfaces (RISs) have the potential to significantly improve the communication quality in the sixth-generation (6G) of wireless networks, by intelligently reconfiguring the wireless propagation of EM signals via low-cost passive reflecting elements (meta-atoms) integrated into planar surfaces. For example, in communication scenarios with obstacles between the transmitter and receiver, virtual line-of-sight links can be created through RIS reflections to improve the desired received signal strength. In this case, the wireless coverage is also extended. Additionally, by configuring the reflection coefficients of RIS elements appropriately, the co-channel/inter-cell interference can be suppressed, EM field exposure can be tamed, and physical-layer security can be further improved.

Although RISs constitute an emerging technology for creating an intelligent wireless radio environment, they are not capable of meeting the demands of future advanced communications applications that involve both communications and computation, such as integrated sensing and/or control with communications. 
In contrast to existing efforts on RISs, in this article, we elaborate on the following significant question: ``\textit{why and how can both intelligent wireless communication and computation be achieved for future 6G?}" Before answering this question, we describe a motivating example. 

\begin{figure*}[t]
  \captionsetup{font={footnotesize}}
\centerline{ \includegraphics[width=5.6in, height=3.0in]{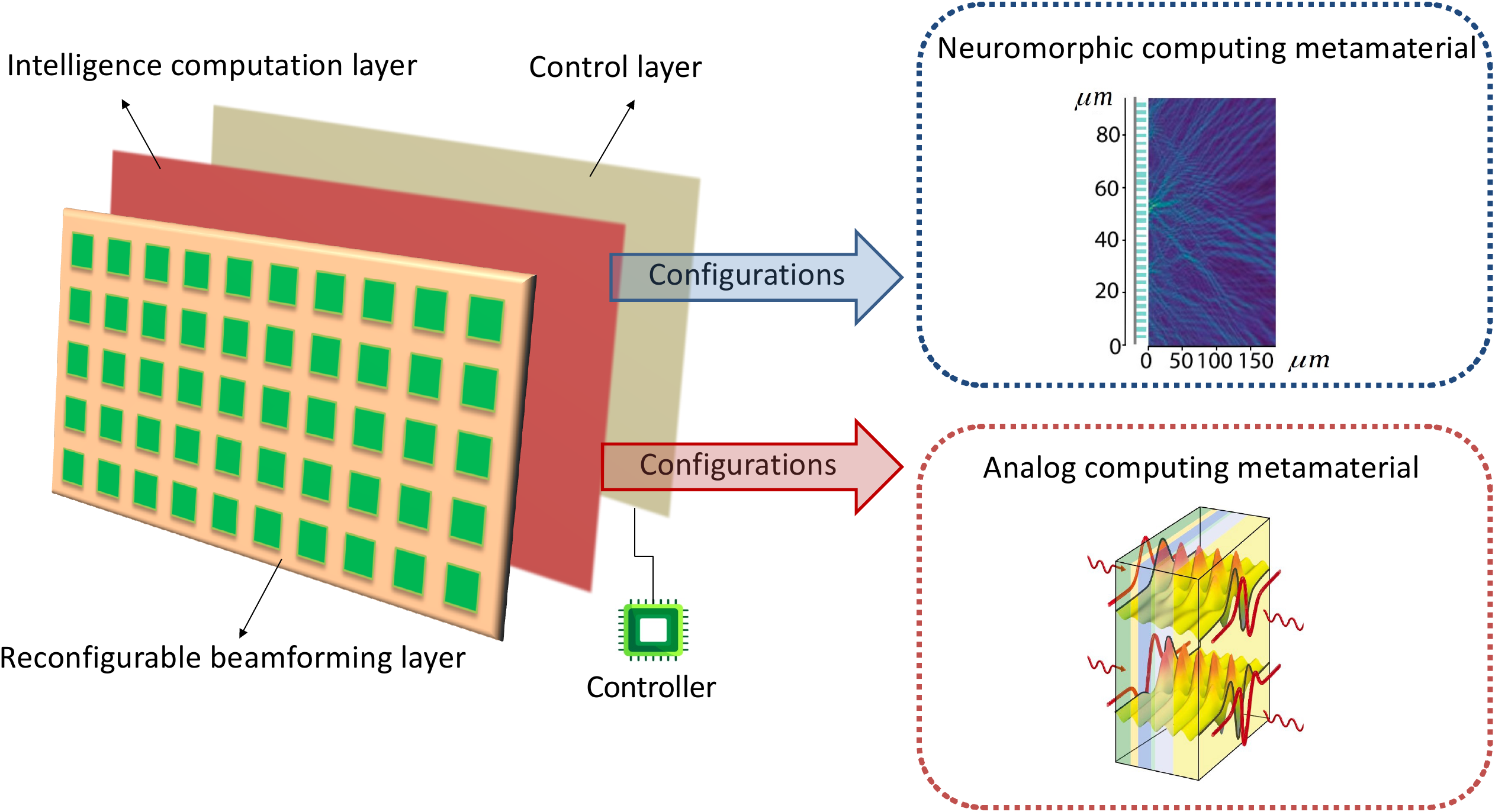}}
\caption{The architecture design of an RICS. It contains three layers: a reconfigurable beamforming layer, an intelligence computation layer, and a control layer. In order to meet the diversification of computational tasks, the intelligence computation layer can be configured by different kind of metamaterials, e.g., neuromorphic computing metamaterials for wireless spectrum learning (shown on the top-right~\cite{Neuromorphic01}) or analog computing metamaterials for secrecy signal processing (shown on the bottom-right~\cite{Science}). 
}
\label{structure}
\end{figure*}

In the conventional RIS-empowered wireless communication systems, the interfering signals tend to dynamically fluctuate and a conventional RIS `blindly' reflects both the desired and interfering signals. In this context, due to the unpredictable nature of interfering signals, undesired reflections via RISs are becoming a critical challenge, which is known to severely degrade the desired signal at the receiver~\cite{YB_TVT}. In contrast, if the conventional RISs were to be empowered with computational capabilities to perform active sensing for interference estimation, such technical challenges could be mitigated. For instance, a programmable artificial intelligence machine structure has been recently proposed to handle various deep learning tasks (such as wave sensing) via manipulating the reflected or transmitted EM waves~\cite{Cui}, this can be achieved with the aid of field-programmable gate arrays (FPGAs).
Instead, by exploring the intrinsic potential of metamaterial-based computing techniques, certain computing operations, including mathematical functions (such as spatial differentiation, integration, and convolution) and artificial neural inference, can be achieved~\cite{2d_material}. Such structures are referred to as computational metamaterials, which specialize in processing computational tasks on signals or images through neuromorphic computing and/or optical analog computing~\cite{{Science,Neuromorphic01}}. \emph{Inspired by the above, it is foreseen that the next generation of intelligent surfaces will integrate computation with communications functions via computational metamaterials.}

 In this paper, we explore a new RIS structure that exploits the natural superiority of computational metamaterials to simultaneous enable dynamically adjustable signal reflections and computational tasks. In particular, we term these structures as `{reconfigurable intelligent computational surfaces}', and abbreviate them as `{RICSs}'. To realize RICSs, computational metamaterials are introduced and connected with the conventional reflective surfaces, thereby attaining the goal of computations and intelligent reflections.

 


\section{Fundamentals of RICS} 
\label{S2}

Different from materials found in nature, the properties of metamaterials (such as permittivity and permeability) stem from the form of their meta-atom design. Recently, the interest in analog computing was revived in the context of metamaterials~\cite{Optical01}. In this context, the proposed RICS belongs to a composite material, which is designed and optimized to function as a tool to control the EM waves as well as to perform computation tasks. 
As conceptually sketched in Fig.~\ref{structure}, an RICS is composed of a smart controller and three layers: the {\textit{reconfigurable beamforming layer}}, the {\textit{intelligence computation layer}}, and the {\textit{control layer}}. The first two multifunctional layers interplay with each other and should be jointly configured. The inner control layer is a control circuit board which is triggered by a smart controller, which focuses on adjusting the tunable parameters of the beamforming layer and can be implemented by a field-programmable gate array. 

In the following, we introduce the design of the first two layers: the reconfigurable beamforming layer and the reconfigurable computation layer. Then, we demonstrate the architectural design of the RICS, which can be used to reflect signals as well as perform computational operations.

\begin{figure*}[t]
  \captionsetup{font={footnotesize}}
\centerline{ \includegraphics[width=7.55in, height=2.4in]{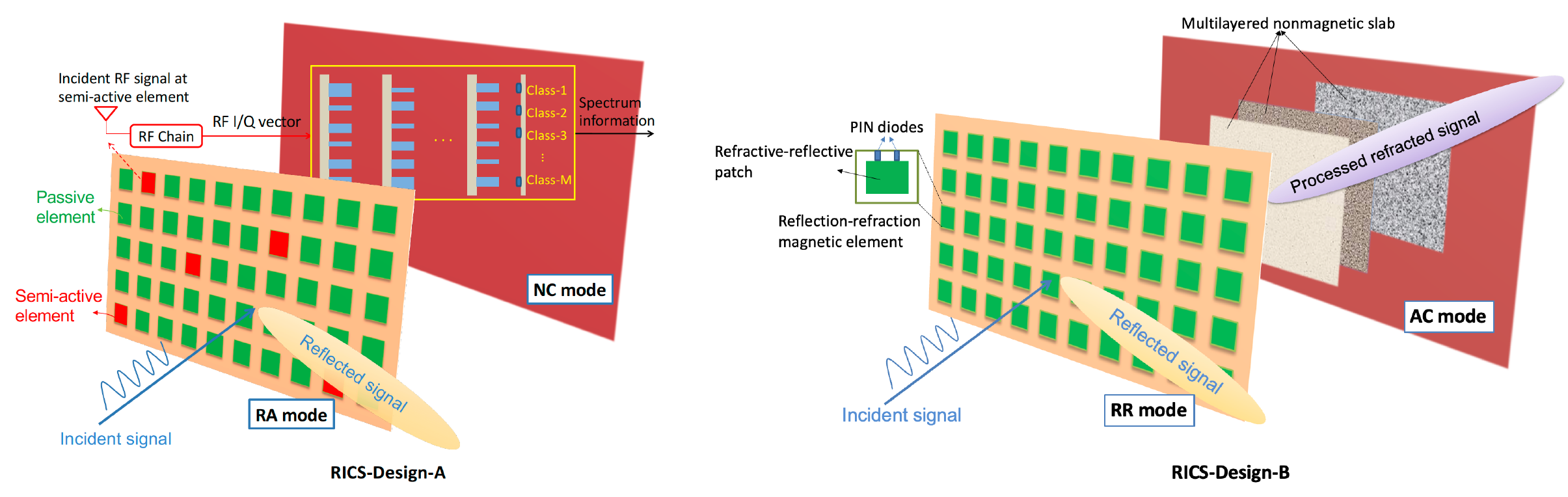}}
\caption{The two kinds of design for RICSs, where the RICS-Design-A is shown on the left side and the RICS-Design-B is shown on the right side.
}
\label{design}
\end{figure*}

\subsection{Reconfigurable Beamforming Layer} 

The reconfigurable beamforming layer commonly comprises a number of tunable elements, which can be dynamically configured to intelligently \textit{reflect}, \textit{refract} or \textit{absorb} the incident radio frequency (RF) signal. In particular, due to the specific computation demand of applications, these tunable elements can be designed to have four kinds of operations:
\begin{itemize}
\item \textbf{Reflection}. This operation indicates that the elements act just like the conventional passive elements that reflect the incident RF signal.

\item \textbf{Refraction}. In this operation, the incident RF signal can be simultaneously reflected and refracted towards both sides of the reconfigurable beamforming layer. For instance, by controlling the ON-OFF state of the positive-intrinsic-negative (PIN) diode of each element~\cite{omni}.

\item \textbf{Absorption}. By enabling this operation, some portion of the tunable elements work as receivers with reception RF chains, thereby allowing to further process the received signal in the digital domain.  

\item \textbf{Storage and retrieval}. This operation enables the storage and retrieval of the EM waves by exploiting the electromagnetically induced transparency effect~\cite{Storage}. 

\end{itemize}

 Based on the configured operations of the tunable elements, the operating mode of the reconfigurable beamforming layer can be categorized into two modes:
\begin{itemize}
\item \textbf{Reflection-absorption (RA) mode}. This mode mainly consists of two types of elements: the conventional reflecting elements and \textit{semi-active elements} for incident RF signal processing. Specifically, for the semi-active RICS elements, only RF front-end, analog-digital converter, and down-conversion are required for obtaining the Inphase-and-Quadrature (I/Q) sequences, thereby achieving some information from the incident wireless signals, e.g., wireless spectrum occupation sensing. Since the baseband processing unit such as signal decoding is not necessary, 
 the deterioration of the signal reflection will not be an issue because only a very few of semi-active elements is required for achieving I/Q samples.


\item \textbf{Reflection-refraction (RR) mode}. To simultaneously realize reflection and refraction, the incident energy is split into two parts: some energy is reflected while the remaining of the energy is refracted to serve users located on the opposite side. 
\end{itemize}

\subsection{Intelligence Computation Layer}

Determined by the growing demands of computational applications, the intelligence computation layer may consist of different kinds of computational metamaterials, such as neuromorphic computing metamaterials and analog computing metamaterials.

\subsubsection{Neuromorphic Computing Metamaterials} To realize artificial neural computing with faster speed and lower energy consumption, optical neuromorphic computing was proposed, especially for achieving {multi-class classification}, by leveraging optical reflection through neuromorphic metamaterials~\cite{Neuromorphic01}, as shown in top-right of Fig.~\ref{structure}. For instance, the neuromorphic computing metamaterial based intelligence computation layer can consist of an array of TiO$_2$ pillars on top of a SiO$_2$ substrate~\cite{structure}.

\textit{Principle:}  Emerging neuromorphic metamaterials generally consist of multiple layers of nanostructures, which are composed of an array of nanoribbons. By changing the size of the ribbons, the amplitude and phase of scattered light can be controlled. Similar to the training of traditional deep neural networks, e.g., the convolutional neural networks, training the neuromorphic metasurface is also a gradient descent process that minimizes a loss function. The difference is that the additional trainable parameters of neuromorphic metasurface include the widths of the nanoribbons. After going through a few layers of the appropriately trained neuromorphic metasurface, the output light becomes a focused beam, which points towards a spatial location corresponding to the inferred class. 

\textit{Operating mode:} The operating mode of the intelligence computation layer via neuromorphic computing metamaterials is denoted as the \textbf{neuromorphic-computing (NC) mode}. For the NC mode, the intelligence computation layer consists of multiple tiers of nanostructures, which are composed of an array of nanoribbons on top of a dielectric substrate. The well-trained intelligence computation layer particularly serves the purpose of classification problem via optical neuromorphic computing. For instance, when a plane wave illuminates an object and passes through the intelligence computation layer, this layer then scatters the light in a way that is equivalent to artificial neural computing. In general, the input of neuromorphic computing metamaterials is the light scattered by an object, which is usually resized and converted into a digital vector first. 

\subsubsection{Analog Computing Metamaterials} 
 In recent years, the study of analog computing through metamaterials has attracted wide attention due to the advantages of parallel processing with ultra-high speed. Different from the fresh literature of intelligent surfaces design that conventionally uses digital units (such as received RF chains and computational/storage) to perform computing tasks, the intelligence computation layer with analog computing metamaterials is able to perform computation tasks (e.g., signal processing),  as highlighted in the bottom-right of Fig.~\ref{structure}. To replace circuits with computing metamaterials, two approaches can be implemented by letting EM waves propagate through metamaterials: the Green's function approach and the metasurface approach~\cite{Optical01}. 
 
\textit{Principle:} Owning to the powerful wave manipulation abilities and subwavelength characteristics, the EM metamaterial can perform mathematical operations, such as spatial integration, differentiation, and convolution. There exist two popular approaches to achieve this functionality: 1) the metasurface approach, and 2) the Green's function (GF) approach, by which the computation can be directly performed on an analog signal. Specifically, the metasurface approach performs signal processing in the Fourier domain based on suitably designed metamaterial blocks that can perform mathematical operations. Each metasurface block is composed of a layered structure of two alternating materials, e.g., Aluminum-doped zinc oxide and silicon. The metasurface approach consists of three sub-blocks: two Fourier transformers via graded-index and lenses, and an optical metasurface between the two graded-indexs for realizing the mathematical operation of choice. 
By suitably manipulating the impinging wave to propagate through the metamaterial blocks, signal processing can be achieved accordingly.
In the GF approach, the multi-layer structure is composed of a stack of subwavelength metamaterial (e.g., dielectric) slabs. By optimizing the permittivity, permeability, and thickness of each slab, it is feasible to carry out the computation directly in a spatial domain without involving additional Fourier lenses.

\textit{Operating mode:} The working mode of the intelligence computation layer via analog computing metamaterials is denoted as the \textbf{analog-computing (AC) mode}. The AC mode particularly serves the purpose of signal processing via mathematical-based analog computing. In particular, the intelligence computation layer that works in this mode consists of multi-tiered dielectric slabs, thereby allowing the synthesis of mathematical operations of interest by realizing the desired GF. By optimizing the permittivity, permeability, and thickness of each slab, the mathematical-operation layer can act as certain mathematical operations on the incident refracted signal to match the considered transfer function without involving additional Fourier lenses.  

%
%

\subsection{RICS Architecture Design}

Based on the configurations of the reconfigurable beamforming layer and the intelligence computation layer, we present two kinds of RICS designs:
\begin{itemize}
\item \textbf{RICS-Design-A: RA+NC.} This design is achieved via neuromorphic computing metamaterials, as shown on the left side of Fig.~\ref{design}. Taking wireless spectrum sensing as an example, to explore the potential of NC mode, data visualization needs to be utilized to map the wireless spectrum data into an unique image, which is, in turn, suitable for the neuromorphic computing metamaterials~\cite{Li_Bo}. 
As illustrated in Fig.~\ref{design}, when the incident RF signal arrives at the reconfigurable beamforming layer working at the RA mode, the impinging signal is received via the semi-active absorption elements, while the remaining reflection elements reflect the signal in a conventional passive way. The received signal is transferred to an I/Q vector, which is then mapped into an image and further fed into the intelligence computation layer working as the NC mode. The final output indicates the class inferring the components of the incident RF signal. Notably, different from the RIS elements design for sensing and reflection~\cite{full_duplex}, semi-active elements are used and only a quadratic computational complexity is involved for inference.


\item \textbf{RICS-Design-B: RR+AC.} This design is achieved via analog computing metamaterials, as shown on the right side of Fig.~\ref{design}. Different from the RICS-Design-A, when the incident RF signal arrives at the reconfigurable beamforming layer, the incident energy is divided into two parts. In this case, the reconfigurable beamforming layer should be configured as the RR mode since the input of the intelligence computation layer is the original analog signal. In particular, some energy is used to reflect the impinging signal and the rest of the energy is for refracting the signal. Then the refracted signal is considered as the input to the intelligence computation layer with the AC mode. By performing analog computing, the output of the intelligence computation layer demonstrates the specific mathematical operation of the incident signal.
\end{itemize} 

To demonstrate the potential of RICSs in wireless communications and how to apply the outcome of an RICS, in the following Section~\ref{S3} and Section~\ref{S4}, we present two illustrative applications of RICSs in wireless communications: intelligent spectrum sensing and secure wireless communications.

\section{Integrated Sensing and Communication via RICS-Design-A} 
\label{S3}

In this section, we present an RICS-Design-A application,  which takes advantage of the neuromorphic computing metamaterials to enable wireless spectrum sensing and communication.

\subsection{Motivation}



In the context of ISAC, wireless spectrum sensing through RF signals has become an important application in future 6G wireless networks. Currently, RIS has been commonly used to improve the quality of wireless links by appropriately reflecting the incident signals. Notably, due to the unpredictable superposition of the wireless signals, undesired reflections of both the desired and interfering signals become a critical challenge, which may cause a deleterious effect on the receiver via the conventional RIS. To address this challenge, it turns out that exploring wireless environments via spectrum sensing becomes beneficial, which, however, requires large amount of computing resources and power. This motivates the necessity to learn and infer the wireless spectrum via RICS-Design-A.


\subsection{Procedure}
 Due to the uniqueness of the wireless signal, the wireless spectrum sensing can be considered as a classification problem, which is addressed via a trained optical neural network (ONN) model in the intelligence computation layer of an RICS. With the inferred spectrum information at RICS, the BS can improve the spectrum efficiency  by allocating the wireless resources intelligently for future 6G networks. 
 
\begin{figure}[t]
  \captionsetup{font={footnotesize}}
\centerline{ \includegraphics[width=3.55in, height=2.2in]{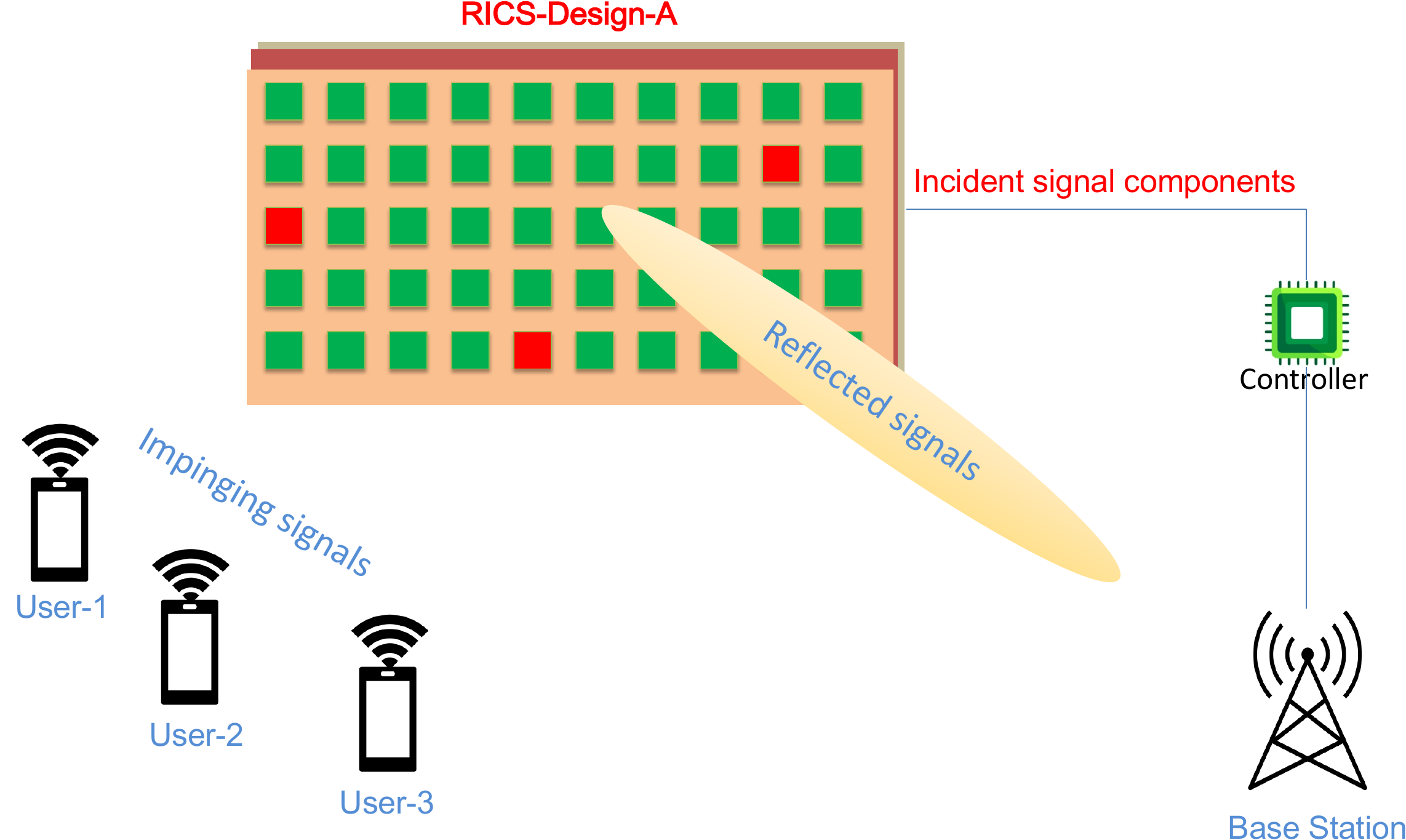}}
\caption{An illuminative example of RICS-aided intelligent spectrum sensing, where three users transmit to a base station via the RICS-Design-A.}
\label{case1}
\end{figure}

An illuminative example is illustrated in Fig.~\ref{case1}, where three users (e.g., $U_1$, $U_2$ and $U_3$) communicate with a base station (BS) via the RICS. The BS maintains a control link with a controller of the RICS, where the RICS sets the configuration as the RICS-Design-A.    In such setups in Fig.~\ref{case1}, an $8$-class classification problem is described as below.
\begin{itemize}
\item Class 1 - `Idle': noise only.
\item Class 2 - `${U_{1}}$': user $U_1$ only.
\item Class 3 - `${U_{2}}$': user $U_2$ only.
\item Class 4 - `${U_{3}}$': user $U_3$ only.
\item Class 5 - `${U_{1}\!+\!U_{2}}$': users $U_1$ and $U_2$.
\item Class 6 - `${U_{1}\!+\!U_{3}}$': users $U_1$ and $U_3$.
\item Class 7 - `${U_{2}\!+\!U_{3}}$': users $U_2$ and $U_3$.
\item Class 8 - `${U_{1}\!+\!U_{2}\!+\!U_{3}}$': users $U_1$, $U_2$ and $U_3$.
\end{itemize}

To realize spectrum learning via RICS-Design-A, we first train an ONN model in the intelligence computation layer. Specifically, we collect RF traces by building a universal software radio peripheral based testbed and store as I/Q sequences. 
With the collected I/Q data with different signal combinations, the ONN model is trained offline via stochastic gradient descent method, 
 which is performed repeatedly until the loss function converges.

After the ONN model is trained appropriately, wireless spectrum sensing can be achieved via online inference at the RICS. Specifically, once the incident RF signals arrive at the reflection-absorption layer, a portion of elements reflect the signal in a conventional way, the remaining elements vectorize and process the signal. Note that before feeding the I/Q vector to the trained intelligent computation layer, data preprocessing (such as frequency adapting and data visualization~\cite{Li_Bo}) is required. Then the spectrum information is output by performing forward calculation via ONN inference\footnote{Different from conventional RISs whose working bandwidth maybe very limited due to their inherent implementation restricts, the RF signals impinging at the RICS are encouraged to be mixed through the same frequency bandwidth, so as to obtain the I/Q samples at the RF chain for further spectrum sensing.}.

\subsection{Illustrative Results}
The considered simulation scenario for evaluating the RICS-Design-A is shown in Fig.~\ref{case1}, where three users send data to the BS from time to time using the transmit power $200$ mW and the data payload size of each user is $1000$ bits. 
The distance between the users and the RICS is 60 $\rm m$, the distance between the RICS and the BS is 80 $\rm m$, the incident angle between the users and the BS is $160\rm{^o}$. Moreover, the noise power density is $-174$ dBm/Hz, the wireless bandwidth is $10$ MHz and the power ratio of the reflected and refracted signals is $1$. 

\begin{figure}[t]
  \captionsetup{font={footnotesize}}
\centerline{ \includegraphics[width=2.85in, height=2.35in]{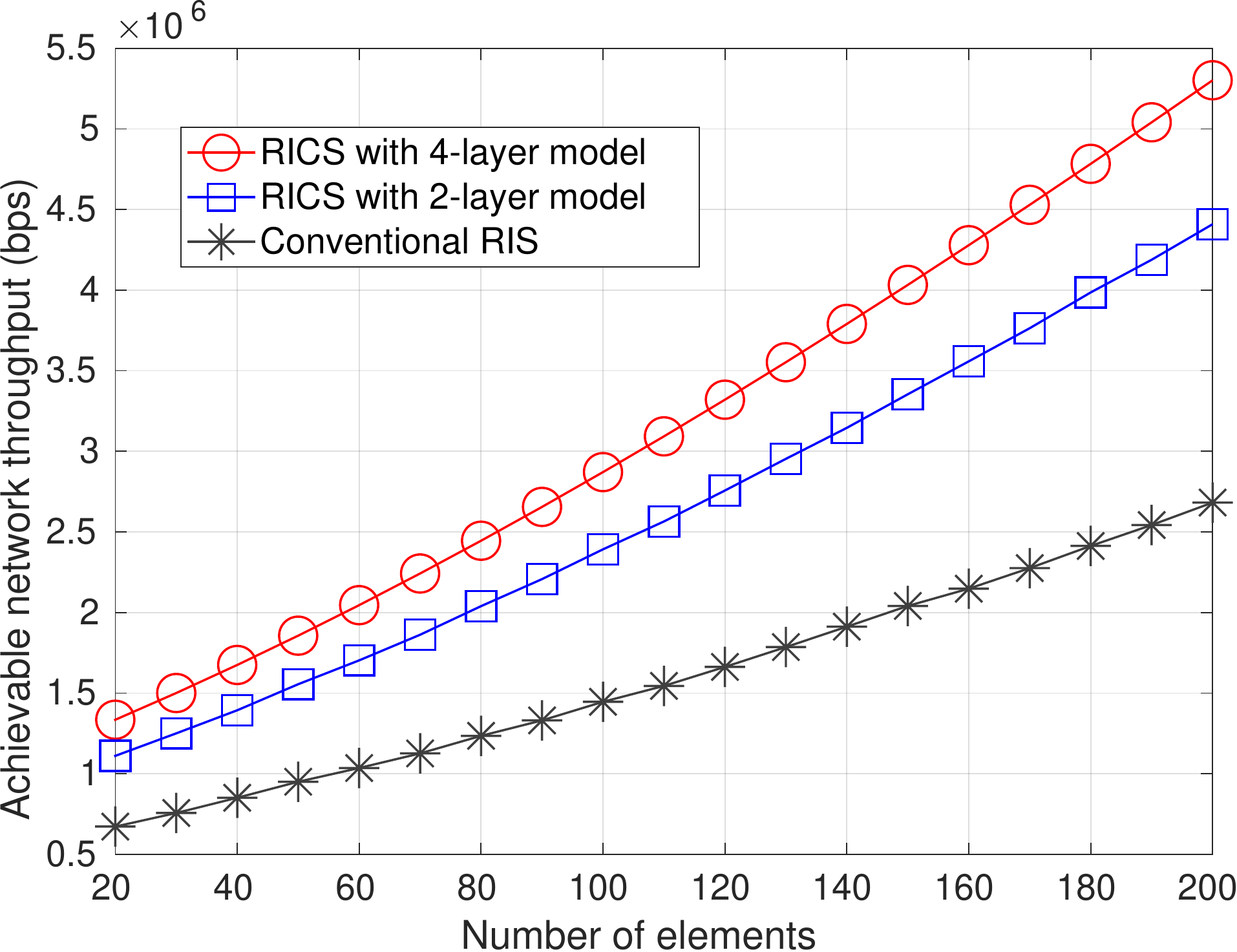}}
\caption{Achieved network throughput versus the number of elements.}
\label{result1}
\end{figure}

We consider demonstrating the wireless spectrum sensing that is achieved via the trained ONN model for classification~\cite{ONN}. Specifically, we trained a 2-layer model and 4-layer model based on the collected RF dataset, which consists of a total of $100$ million I and Q samples indicating one of the eight classes. After the models converge, the 2-layer model can achieve $85\%$  accuracy for spectrum sensing and the 4-layer model can reach $90\%$. This demonstrates that the inference accuracy can be improved when more layers are being used. 

To evaluate throughput performance, we implement a time division multiple access scheme for RICS and RIS, respectively. Then we evaluate and compare the achieved network throughput, as illustrated in Fig.~\ref{result1}. Different from statistical time slots allocation with the conventional RIS, we observe that the RICS transmission schemes are always superior to the conventional RIS scheme since the RICS can infer the incident signal components, thereby enabling the BS to allocate time slots for the active users intelligently. Reflected in Fig.~\ref{result1}, we also note that the inference accuracy achieved by the trained ONN model at the RICS affects the network throughput significantly. In particular, compared to the RICS with 2-layer model, the RICS with 4-layer model outperforms and the performance gap becomes larger as the number of elements increases. 


\section{Secure Wireless Communications via RICS-Design-B} \label{S4}
In this section, we present an RICS-Design-B application, which takes advantage of the analog computing metamaterials to achieve secure wireless communications.

\subsection{Motivation}
In recent years, physical layer security has attracted increasing attention from research and industrial communities.  Suppose the existence of an eavesdropper in the network. When the user transmits data to the legitimate receiver, the information leakage issue may occur since the eavesdropper also can receive the wireless signal that comes from this considered user. To address this challenge, deploying a conventional RIS for generating a tuned reflected signal has become a recognized solution\cite{add}. However, the unavailability of computational capability and prior information on the eavesdroppers has become a stumbling block for reducing information leakage via the RIS in practice, especially when the eavesdroppers located on the \textit{opposite-side} of the RIS. This motivates the necessity to process the incident signal via RICS-Design-B, thereby suppressing the received signal-to-interference-plus-noise at the eavesdroppers. 

\begin{figure}[t]
  \captionsetup{font={footnotesize}}
\centerline{ \includegraphics[width=3.25in, height=2.4in]{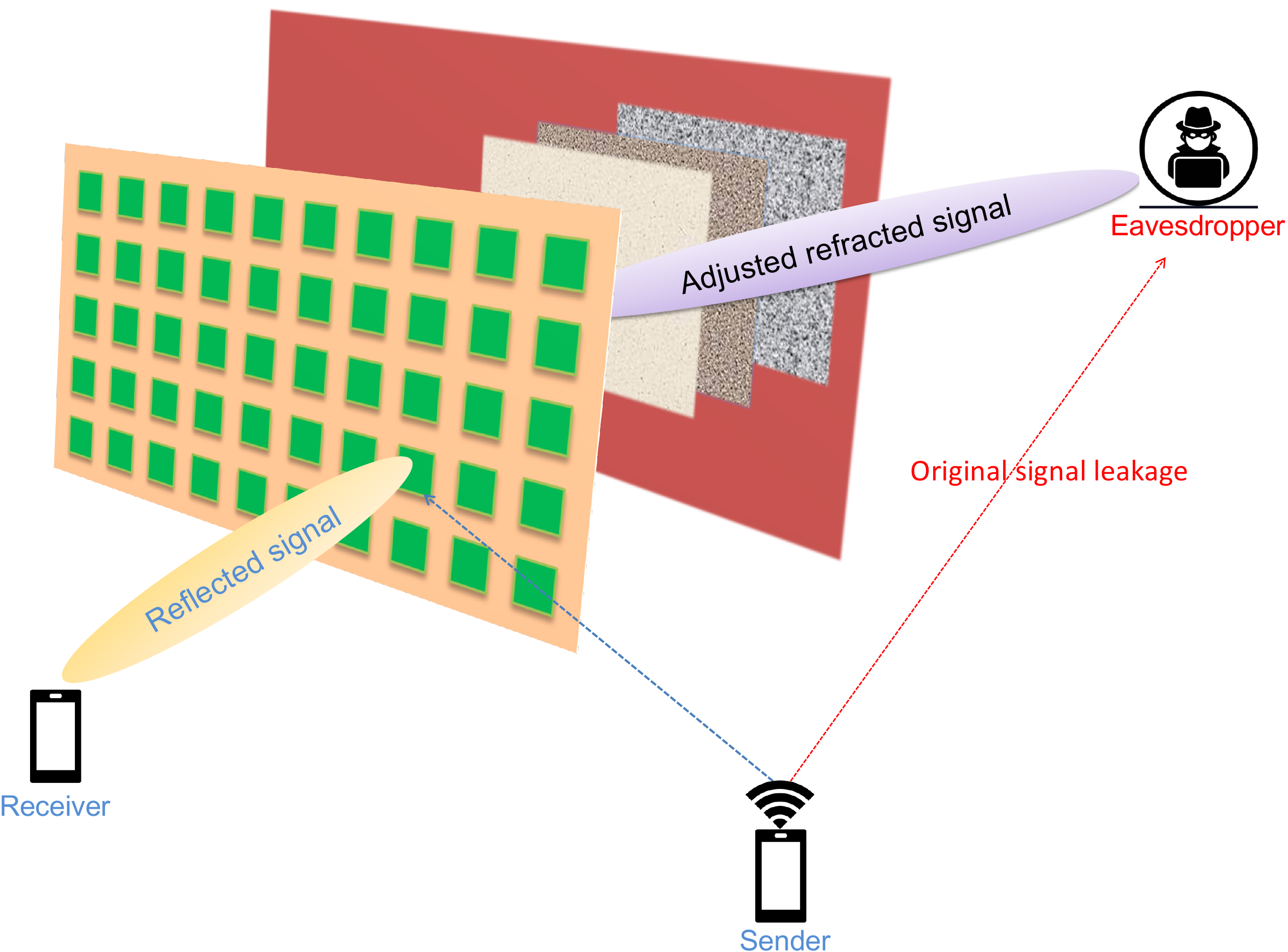}}
\caption{An illustrative example of secure wireless communication, where   an intended interfering signal is generated via the RICS-Design-B to worsen the quality of the signal at the eavesdroppers.}
\label{case2}
\end{figure}

\subsection{Procedure}
With the implementation of an RICS, the incident signal can be refracted at the reconfigurable beamforming layer and adjusted appropriately by the intelligence computation layer. Then the processed signal can be destructively added with the non-reflected signal at the eavesdropper to neutralize the leaked signal.

 An illustrative example is shown in Fig.~\ref{case2}, where a sender transmits data to a receiver and an eavesdropper is nearby. We note that the wireless signal that comes from the sender can also be received by the eavesdropper.  Different from the conventional RIS without computing capability, the RICS-Design-B can be exploited to generate an intended interfering signal to worsen the quality of the leakage of the signal at the eavesdroppers.
Specifically, when the incident signal arrives at the reconfigurable beamforming layer of the RICS, the energy of the incident signal is divided into two parts. Some of the incident energy is reflected to serve the desired receiver located on the same side as the sender, and the rest of the energy works for impinging signal refraction\footnote{Since the channel model of the reflected and refracted signals may not be symmetric, the power ratio of the reflected and refracted signals in the reconfigurable beamforming layer could be appropriately optimized~\cite{omni}, thereby providing an extra degree of freedom for enhancing the RICS-aided communication performance.}.

Then, by feeding the refracted signal to the intelligence computation layer that works in the AC mode, an intended interfering signal can be appropriately generated by performing the mathematical operation, e.g., frequency shifting, to worsen the leaked signal at the eavesdroppers. Therefore, instead of relying on higher-layer encryption, the RICS enables the exchange of confidential messages over a wireless medium in the presence of unauthorized eavesdroppers.



\begin{figure}[t]
  \captionsetup{font={footnotesize}}
\centerline{ \includegraphics[width=2.85in, height=2.25in]{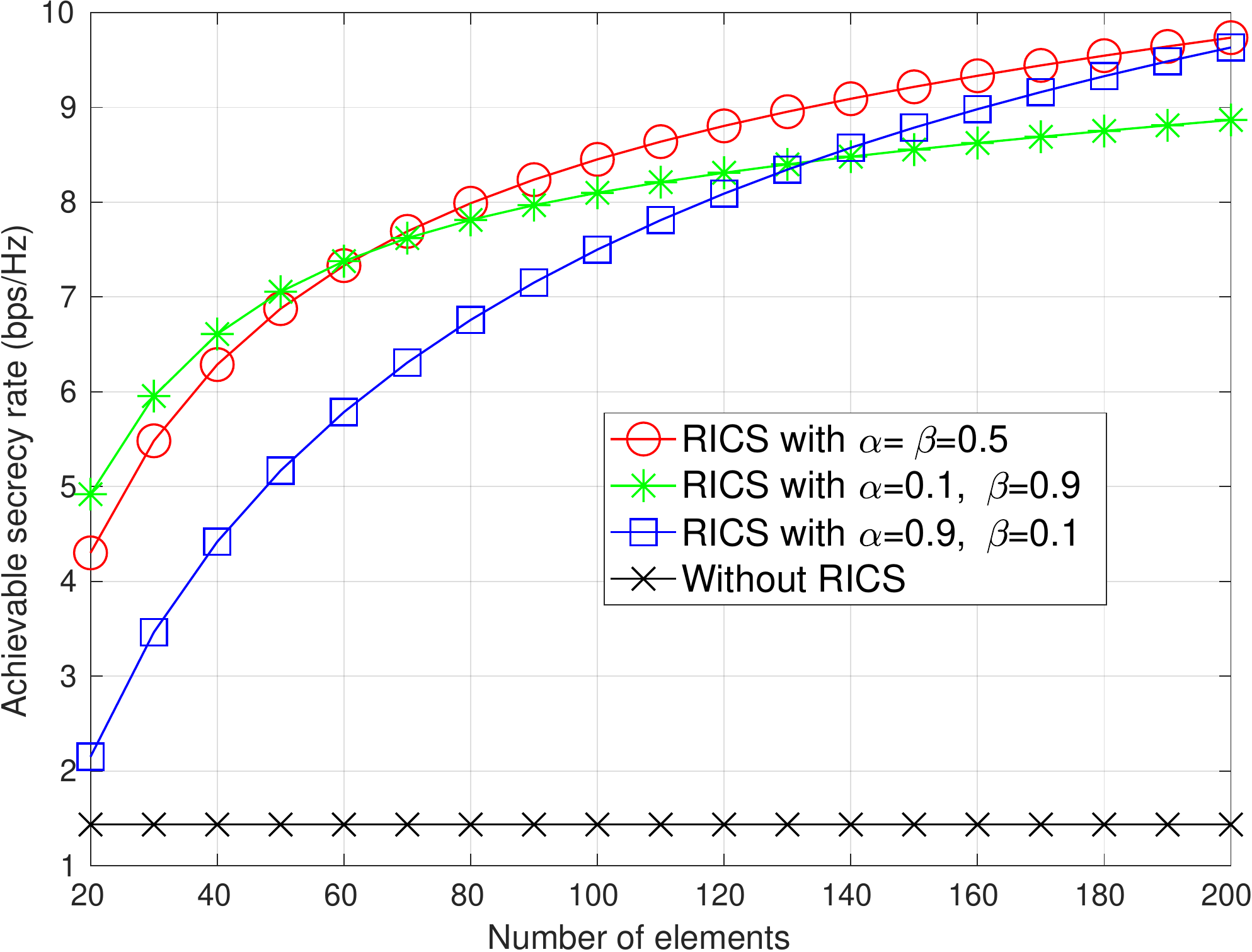}}
\caption{Achievable secrecy rate versus the number of elements.}
\label{result2}
\end{figure}

\subsection{Illustrative Results}
The considered simulation scenario for evaluating the RICS-Design-A is shown in Fig.~\ref{case2}, where the distance between the RICS and the eavesdropper is 50 $\rm m$ and other critical parameters keep unchanged. The power ratio between the reflected signal and the incident signal is denoted as $\alpha$, the power ratio of between the refracted signal and the incident signal is denoted as $\beta$, and $\alpha+\beta=1$ holds in the simulations. 
To demonstrate the performance of secure wireless communication via the RICS-Design-B, we evaluate the achievable secrecy rate in bits/second/Hertz (bps/Hz), which can be expressed as the achievable secrecy rate which is the difference between the achievable rate of the legitimate link and the eavesdropper link. 


Fig.~\ref{result2} compares the achievable secrecy rates of four schemes. We observe that the achievable secrecy rate of the three schemes that are based on the RICS is higher than that of the scheme `Without RICS', and the performance gap increases as the number of RIS elements grows. This indicates that with more reflecting elements, the reflect and refract beamforming design of the RICS becomes more effective, thereby achieving higher gains of secrecy rate. We also observe that the power ratio coefficients, $\alpha$ and $\beta$, play a significant impact on the achievable secrecy rate performance. In particular, when the number of elements is small, e.g., less than $60$,  a lower value of $\alpha$ brings a beneficial impact on the RICS. As the number of RIS elements increases, the scheme with a larger value of $\alpha$ outperforms.
It is worth mentioning that there exists an optimal trade-off between the reflected power and the refracted power for a given number of elements of the RICS.

\section{Research Challenges and Directions} \label{S5}
Taking advantage of metasurface technologies, computational metamaterials provide potential solutions for the design of next-generation intelligent surfaces in future 6G networks. 
Despite the two motivating examples above, there exist more promising application scenarios that integrate communication with computing for the proposed RICSs, such as artificial intelligence (AI) based localization with fingerprinting, interference suppression for the cell edge users, and so on.
However, to meet urgent computing demands, the ability of RICSs for carrying out computational tasks gives rise to  several research challenges.


\subsection{Nonlinear Computing Design}
Note that the analog computational metamaterials discussed in the proposed RICS structure are usually related to linear functionalities. With the emerging demands of complex applications, investigating the possibility of performing nonlinear processing operations through computational metamaterials is necessary. In this event, nonlinear signal processing technologies will become a potential player for improving the performance of future 6G wireless system. 

To address this challenge, the concept of metasurfaces is undergoing transformation to the nonlinear regime for the realization of nonlinear functionalities~\cite{Optical01}. 
In such a context, investigating the possibility of performing more complex calculations and operations via RICSs with nonlinear-enabled computational metamaterials becomes an attractive research direction.

\subsection{Multi-functional Computing Design}
At the metamaterials level, computation efficiency of metamaterials plays an important role, and enhancing the speed of computation
is important for the implementation of RICSs. 
Compared to the single-task processing for analog computing design, a promising trend in computational metamaterials is multifunctional wave-based analog computing, in which several computational tasks are performed simultaneously via different independent processing channels~\cite{Parallel}. As a result, such multifunctional computing metamaterials can provide the unique possibility of parallel processing of information and enhance the computation efficiency substantially. Under this trend, the development of parallel computational metamaterials will open a new route for the future intelligent metasurfaces design with accelerated signal processing capability.

\subsection{Artificial Intelligence Driven Design}
The RICS configurable profile (such as phase shift matrices in the reconfigurable beamforming layer and settings of the intelligence computation layer) has to be optimized for enhancing the system performance.
In practical deployments, each RICS could be equipped with hundreds of meta-atoms. Most of the existing contributions tend to rely on mathematical model-based optimization methods, which generally require a large number of iterations to find a near-optimal solution due to the non-convex natures of the constraints and the objective function.

With the development of AI technologies, investigating the possibility of intelligent surface design by use of deep neural networks or other machine learning structures is of interest.
Compared with the conventional optimization methods for non-convex equation solving, deep learning could help to quickly generate optimal solutions for the configurations of intelligent surfaces with higher accuracy~\cite{YB_TMC}. On the far horizon, we may imagine combining computational metamaterials with machine learning to enable the realization of AI-driven RICS design.

\section{Conclusion} \label{S6}
In this article, we have introduced a novel concept of RICSs for meeting emerging demands on wireless communication and computation. In contrast to purely reflective RISs, RICSs are capable of reflecting and processing impinging signals based on computational metamaterials. We have presented two representative RICS designs and discussed their feasibility. Preliminary results are shown to demonstrate the advanced features of RICS. We finally highlight several research challenges, which pave the way for realizing the potential of RICSs for future 6G wireless networks.


\appendices




\ifCLASSOPTIONcaptionsoff
  \newpage
\fi


\begin{thebibliography}{00}

\bibitem{YB_TVT}
B. Yang, X. Cao, C. Huang, C. Yuen, L. Qian, and M. Di Renzo, ``Intelligent spectrum learning for wireless networks with reconfigurable intelligent surfaces," \textit{IEEE Trans. Veh. Technol.}, vol. 70, no. 4, pp. 3920-3925, Apr. 2021.

\bibitem{Cui}
G. C. Alexandropoulos, K. Stylianopoulos, C. Huang, C. Yuen, M. Bennis, and M. Debbah, ``Pervasive machine learning for smart radio environments enabled by reconfigurable intelligent surfaces," \textit{IEEE Proceedings}, to appear, 2022.

\bibitem{2d_material}
C. Liu, H. Chen, S. Wang, Q. Liu, Y.G. Jiang, D.W. Zhang, M. Liu, and P. Zhou, ``Two-dimensional materials for next-generation computing technologies," \textit{Nature Nanotechnology}, vol. 15, no. 7, pp. 545-557, Jul. 2020.

\bibitem{Neuromorphic01}
Z. Wu, M. Zhou, E. Khoram, B. Liu, and Z. Yu, ``Neuromorphic metasurface," \textit{Photonics Research}, vol. 8, no. 1, pp. 46-50, Jan. 2020.

\bibitem{Science}
A. Silva, F. Monticone, G. Castaldi, V. Galdi, A. Al\`u, and N. Engheta,  ``Performing mathematical operations with metamaterials," \textit{Science}, vol. 343. no. 6167, pp. 160-163, Jan. 2014.

\bibitem{Optical01}
F. Zangeneh-Nejad, D.L. Sounas, A. Al\`u, and R. Fleury, ``Analogue computing with metamaterials," \textit{Nature Reviews Materials}, vol. 6, pp. 207-225, Mar. 2021.


\bibitem{omni}

S. Zhang, H. Zhang, B. Di, Y. Tan, M. Di Renzo, Z. Han, 
H. Vincent Poor, and L. Song, ``Intelligent omni-surfaces: ubiquitous wireless transmission by reflective-refractive metasurfaces," \textit{IEEE Trans. Wireless Commun.}, vol. 21, no. 1, pp. 219-233, Jan. 2022.


\bibitem{Storage}
T. Nakanishi and M. Kitano, ``Storage and retrieval of electromagnetic waves using electromagnetically induced transparency in a nonlinear metamaterial," \textit{Applied Physics Letters}, vol. 112, pp. 1-5, May. 2018.





\bibitem{structure}
M. Khorasaninejad, Z. Shi, A. Y. Zhu, W. T. Chen, V. Sanjeev, A. Zaidi,  and F. Capasso, ``Achromatic metalens over 60 nm bandwidth in the visible and metalens with reverse chromatic dispersion," \textit{Nano Letters}, vol. 17, no. 3, pp. 1819-1824, Jan. 2017.

\bibitem{Li_Bo}
G. Feng, B. Li, M. Yang, and Z. Yan, ``V-CNN: data visualizing based convolutional neural network," \textit{IEEE Int. Conf. Signal Process., Commun., Comp.}, Sep. 2018, Qingdao, China.

\bibitem{full_duplex}
G. C. Alexandropoulos, N. Shlezinger, I. Alamzadeh, M. F. Imani, H. Zhang, and Y. C. Eldar, ``Hybrid reconfigurable intelligent metasurfaces: enabling simultaneous tunable reflections and sensing for 6G wireless communications," \textit{arXiv preprint arXiv:2104.04690}, Apr. 2021.


\bibitem{ONN}
I. A. Williamson, T. W. Hughes, M. Minkov, B. Bartlett, S. Pai, and S. Fan,  ``Reprogrammable electro-optic nonlinear activation functions for optical neural networks," \textit{IEEE J. Sel. Topics Quantum Electron.}, vol. 26, no. 1, pp. 1-12, Feb. 2020.

\bibitem{add}
G. C. Alexandropoulos, K. Katsanos, M. Wen, and D. B. da Costa, ``Safeguarding MIMO communications with reconfigurable metasurfaces and artificial noise," in Proc. of \textit{IEEE Int. Conf. Commun.}, Montreal, Canada, Jun. 2021.

\bibitem{Parallel}
 A. Abdolali, A. Momeni, H. Rajabalipanah, and K. Achouri, ``Parallel integro-differential equation solving via multi-channel reciprocal bianisotropic metasurface augmented by normal susceptibilities," \textit{New Journal of Physics}, vol. 21, 113048, Nov. 2019.

\bibitem{YB_TMC}
X. Cao, B. Yang, C. Huang, C. Yuen, M. Di Renzo, D. Niyato, and Z. Han,  ``Reconfigurable intelligent surface-assisted aerial-terrestrial communications via multi-task learning," \textit{IEEE J. Sel. Areas Commun.}, vol. 39, no. 10, pp. 3035-3050, Oct. 2021.



\end{thebibliography}
\end{document}